\documentclass[twocolumn,showpacs,preprintnumbers,amsmath,amssymb]{revtex4}

\usepackage{graphicx}
\usepackage{dcolumn}
\usepackage{bm}

\begin{document}

\title{Some fundamental problems for an energy conserving adaptive resolution molecular dynamics scheme}
\author{Luigi Delle Site}
\email{dellsite@mpip-mainz.mpg.de}
\affiliation{%
Max-Planck-Institut f\"ur Polymerforschung, Ackermannweg 10, D-55128 Mainz, Germany
}%

\begin{abstract}
Adaptive resolution molecular dynamics (MD) schemes allow for changing the number of degrees of freedom on the fly and preserve the free exchange of particles between regions of different resolution. There are two main alternatives on how to design the algorithm to switch resolution using auxiliary ''switching'' functions; force based and potential energy based approach. In this work we show that, in the framework of classical MD, the latter presents fundamental conceptual problems which make unlikely, if not impossible, the derivation of a robust algorithm based on the potential energy.

\end{abstract}
\maketitle
\subsection{Introduction}
Multiscale modeling and simulation in condensed matter is a field of continuous expansion as the basic properties of an increasing number of systems, relevant to current research, are discovered to strongly depend on a delicate scales' interplay. The massive progress of computer technology together with  the parallel development of novel powerful simulation methods has strongly contributed to this expansion so that by now detailed sequential studies from the electronic scale to the mesoscopic and continuum are routinely performed. However among all these methods of a particular interest are those which deal in a more direct way with the multiscale idea. Typically these schemes are based on a single computational approach which links two or more interconnected scales. One example is the technique used to study  edge dislocation in metals, where local chemistry affects large scale material properties \cite{kax}, or the crack of materials where the rupture of a local interatomic bond is then propagated to the larger scale and again back to the next interatomic bond and so on \cite{Rottler:2002,Csanyi:2004} ; in this case quantum based methods are interfaced with classical atomistic and continuum models within a single computational scheme. A further example is the Quantum Mechanics/Molecular Mechanics scheme \cite{Laio:2002}. This is mainly used for soft matter systems where a fixed region of space requires quantum resolution and the external part is treated at classical atomistic level. Examples are solvation of large molecules where the chemistry happens locally (quantum region) while the statistical effect of the fluctuating environment (solvent) far from the molecules can be treated in a rather efficient way at classical level. In the same fashion there are several more examples (see e.g. Refs.\cite{Jiang:2004,Lu:2005}). All of these anyway are characterized by a non trivial limitation, i.e. the region of resolution is fixed and free exchange of particles with the other regions are not allowed.
While this may not be a crucial point for system involving rigid structures,certainly is a very strong limitation for highly fluctuating systems.
The natural next step to overcome this problem is the design of novel adaptive resolution methods which indeed allow for the exchange of particles among regions of different resolution. In general, in such a scheme a molecule moving from a high resolution region to a lower one, would gradually loose some degrees of freedom until the lower resolution is reached but yet the statistical equilibrium among the two different regions is kept at any instant. Recently some schemes based on this idea have been presented in the literature \cite{jcp,pre1,ensing}. They differ in the way the different resolutions are coupled in the MD algorithm.
The coupling can be achieved either through the potential, slowly passing from an atomistic to a corresponding coarse grained potential (and vice versa), or through the forces, that is slowly passing from a force derived from an atomistic potential to a force derived from the corresponding coarse grained potential (and vice versa). The passage from the atomistic to the coarse grained is controlled by a smooth ''switching function'' which is used to interpolate the two quantities.
For the force based scheme it is not possible to define a potential energy from the interpolation formula, but on the basis of physical arguments this problems can be circumvented \cite{pre2,jpa} as it will be briefly discussed later on; for the potential based scheme obviously the definition of potential energy is the central point. In this sense, the potential based scheme would seem more appealing, however the subject of this work is to show that on the basis of a mathematically rigorous derivation, this scheme is not applicable.
Here we construct the most general adaptive scheme based on the potential and derive the necessary conditions by which one can obtain the switching functions. As an outcome we show that the resulting set of partial differential equations has got boundary conditions such that the system is overdetermined and thus solutions may exists only for trivial cases. Moreover, even in case a solution may exist, further technical problems, due to the nature of the differential equation, arise which make this scheme rather unpractical.
The paper is organized as follows; in the next section a short overview of the force based method is presented, it summarizes its crucial point and enumerates the latest applications. Next the potential based scheme is presented with its general features. Finally, a  general interpolation scheme is used to derive the equation that defines the switching functions. The paper is closed by the discussion and conclusions. 
\subsection{Force based scheme: A short overview of the AdResS method}
According to the previous discussion, a method which has turned to be rather robust is the Adaptive Resolution Simulation (AdResS) \cite{jcp,pre1}. It is based on coupling the atomistic and the mesoscale through an interpolation formula for the atomistic and coarse grained force. At this point it should be mentioned that this approach, as well as the calculations performed in this work, are valid, so far, under the assumption of pair interactions; however due to the large use of pair potentials in atomistic simulation, the AdResS method, as well as the result of this work are nevertheless of interest to the simulation community. 
\begin{figure}[!ht]
\includegraphics[width=5.0cm]{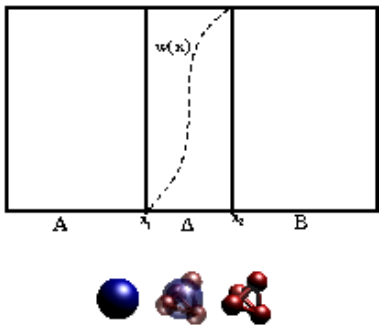}
\caption{\label{fig1} (Color on line) Schematic picture of the partitioning of space in high resolution (atomistic) region ${\bf B}$, low resolution (coarse grained) region ${\bf A}$ and transition region ${\Delta}$. $w(x)$ is the switching function which allows a smooth transition from a coarse grained to an atomistic resolution and vice versa. Below the pictorial representation of a tetrahedron molecule that change resolution according to the position in space is presented. This representation is taken 
from \cite{jcp}}
\end{figure}
Briefly, the space is divided in two regions as for example in Fig.\ref{fig1} a high resolution region, let us call it ${\bf B}$ where the molecule has atomistic resolution and a region ${\bf A}$ where the molecule is coarse grained. In between there is a region ${\bf \Delta}$ where a smooth transition from one resolution to another takes place via a continuous ''switching'' function $w(x)$, such that $w(x_{1})=0; w(x_{2})=1$. The interpolation formula then reads \cite{jcp}:
\begin{equation}
 {\bf F}_{\alpha \beta}=w(X_\alpha)w(X_\beta){\bf F}_{\alpha\beta}^{atom}+[1-w(X_\alpha)w(X_\beta)]{\bf F}^{cm}_{\alpha\beta}
\label{intforce}
\end{equation}
where $\alpha$ and $\beta$ labels two distinct molecules, ${\bf F}_{\alpha\beta}^{atom}$ is derived by the atomistic potential where each atom of molecule $\alpha$ interacts with each atom of molecule $\beta$, and ${\bf F}^{cm}_{\alpha\beta}$ is obtained from  an effective
pair potential between the centers of mass of the coarse-grained molecules; the latter is derived on the basis of the
reference all-atom system. Eq.\ref{intforce}, does not allow to define a potential in the switching region \cite{pre2,jpa}, however, in this scheme such a definition is not required. Actually, all one needs to know is based on the following arguments: the change of resolution can be interpreted in terms of similarity with a geometrically induced first order phase transition with an associated latent heat \cite{pre2}. This interpretation justifies the use of a thermostat, during an MD simulation, in the switching region ${\bf \Delta}$ so that the physical equilibrium is kept. Numerical calculations and applications to rather different systems have shown that indeed this approach is satisfactory (see the applications to a liquid of tetrahedral molecules \cite{pre1}, a polymer solvated in it, \cite{jcp2} and liquid water \cite{jpcm}). The crucial point of this scheme is that Eq.\ref{intforce} is only an ansatz based on satisfying the third Newton law and on numerical simplicity; however the numerical results show that the method indeed gives the correct answers when compared with all atom simulations and its physical interpretation is consistent with the basic principles of equilibrium in statistical mechanics \cite{jpa}. However, one may naturally ask whether the same or a similar interpolation scheme can be applied to potentials and thus preserve the energy conservation as suggested by Ensing {\it et al.} \cite{ensing}. In the following section we show that to build an interpolation scheme similar to Eq.\ref{intforce} but applied to potentials instead of forces {\bf it is not} possible.      
\subsection{A generic scheme based on the potentials}
Let us define two generic switching functions:
$f(X_{\alpha},X_{\beta})$ continuous and differentiable in ${\bf \Delta}$, and outside ${\bf \Delta}$ defined such that:
\begin{eqnarray}
f(X_{\alpha},X_{\beta})=0; X_{\alpha}\ge x_{2}, and, X_{\beta}\ge x_{2}\nonumber\\f(X_{\alpha},X_{\beta})=1; X_{\alpha}, or, X_{\beta} \le x_{1}
\label{bc1}
\end{eqnarray}
and $g(X_{\alpha},X_{\beta})$ continuous and differentiable in ${\bf \Delta}$ and outside ${\bf \Delta}$ defined such that:
\begin{eqnarray}
g(X_{\alpha},X_{\beta})=0; X_{\alpha}\le x_{1} or X_{\beta}\le x_{1}\nonumber\\g(X_{\alpha},X_{\beta})=1; X_{\alpha} and X_{\beta} \ge x_{2}.
\label{bc2}
\end{eqnarray}
Here $X_{\alpha}$ and $X_{\beta}$ are the coordinates along the ${\hat x}$ direction, as represented in Fig.\ref{fig1}, of the center of mass respectively of the generic molecule $\alpha$ and $\beta$.
A generalization of Eq.\ref{intforce} to the potentials using these two generic switching function $f(x)$ and $g(x)$ writes:
\begin{equation}
U^{coupling}=f(X_{\alpha},X_{\beta})U_{cg}+g(X_{\alpha},X_{\beta})U_{atom} 
\label{potint}
\end{equation}
where $U_{coupling}$ is the potential coupling the two resolutions, $U_{cg}=U_{cg}({\bf R}_{\alpha},{\bf R}_{\beta})$ is the coarse grained potential and ${\bf R}_{\alpha},{\bf R}_{\beta}$ the coordinates of the centers of mass; $U_{atom}=U_{atom}({\bf r}_{\alpha i},{\bf r}_{\beta j})$ is the atomistic potential between atom $i$ of molecule $\alpha$ and atom $j$ of molecule $\beta$. Eq.\ref{potint} couples the different scales similarly to what is done by Eq.\ref{intforce} but the with the hypothetical advantage of automatically conserving energy. 
At this point to do molecular dynamics, we need to derive the forces. The following situations are clear: if the molecules are located both in region ${\bf A}$ (coarse grained force), or both in region ${\bf B}$ (atomistic force), or one in ${\bf A}$ and one in ${\bf B}$ (coarse grained force), or one in ${\bf \Delta}$ and one in ${\bf A}$ (coarse grained force). However once the molecules are both in ${\bf \Delta}$ or one in ${\bf \Delta}$ and one in ${\bf B}$, the force must be derived by the whole expression of Eq.\ref{potint}.
Let us calculate the coupling force acting on ${\bf R}_{\alpha}$ and ${\bf R}_{\beta}$. One should keep in mind that ${\bf R}_{\alpha}=\sum_{i=1,n} {\bf r}_{\alpha i}/n$ and equivalently ${\bf R}_{\beta}=\sum_{i=1,n},{\bf r}_{\beta i}/n$, where for simplicity the molecules where chosen to have both $n$ identical atoms. It follows that the force acting on the center of mass  of molecule $\alpha$ is :
\begin{equation}
{\bf F}^{coupling}_{{\bf R}_{\alpha}}=-\frac{\partial U^{coupling}}{\partial {\bf R}_{\alpha}}
\label{fsh}
\end{equation}
which in explicit form writes:
\begin{eqnarray}
{\bf F}^{coupling}_{{\bf R}_{\alpha}}=-f(X_{\alpha},X_{\beta})\frac{\partial U_{cg}}{\partial {\bf R}_{\alpha}}-g(X_{\alpha},X_{\beta})\frac{\partial U_{atom}}{\partial {\bf R}_{\alpha}}-\nonumber\\U_{cg}\frac{\partial f(X_{\alpha},X_{\beta})}{\partial {\bf R}_{\alpha}}-U_{atom}\frac{\partial g(X_{\alpha},X_{\beta})}{\partial {\bf R}_{\alpha}}
\label{leq1}
\end{eqnarray}
taking into account that $\frac{\partial {\bf R}_{\alpha}}{\partial X_{\alpha}}=1$, Eq.\ref{leq1} can be rewritten as:
\begin{equation}
{\bf F}^{coupling}_{{\bf R}_{\alpha}}=f(X_{\alpha},X_{\beta}){\bf F}_{cg}+g(X_{\alpha},X_{\beta}){\bf F}^{atom}_{cm}+{\bf F}_{drift}
\label{leq2}
\end{equation}
where 
\begin{equation}
{\bf F}_{drift}=-U_{cg}\frac{\partial f(X_{\alpha},X_{\beta})}{\partial X_{\alpha}}-U_{atom}\frac{\partial g(X_{\alpha},X_{\beta})}{\partial X_{\alpha}}
\label{fd}
\end{equation}
is a spurious force with no physical meaning which emerges as a consequence of the presence of the switching functions. 
In fact in this case the center of mass of a molecule receives an additional acceleration in the switching region, due to the switching function, which should not be there because from the physical point of view the molecules in any resolution regime {\bf must be equivalent} and the switching function is not a physical quantity. 
The effect that it will have is of drifting particles along the ${\hat x}$ direction.
At this point, the condition to have a well based physical treatment of the particles without artifacts in the dynamics due to the introduction of the switching functions is to recover from Eq.\ref{leq2}, in a mathematical way, the force coupling scheme involving only the atomistic and coarse grained force. This will allow us to determine $f(X_{\alpha},X_{\beta})$ and $g(X_{\alpha},X_{\beta})$ for which the energy is conserved and to have an algorithm that in principle works rather well as shown by the AdResS scheme. The physical condition to do so, as implicitly suggested  by Ensing {\it et al.}, is:
\begin{equation}
 {\bf F}_{drift}=0
\end{equation}
and in this case, translated into the mathematical condition, becomes:
\begin{equation}
U_{cg}\frac{\partial f(X_{\alpha},X_{\beta})}{\partial X_{\alpha}}+U_{atom}\frac{\partial g(X_{\alpha},X_{\beta})}{\partial X_{\alpha}}=0
\label{partdfiff1}
\end{equation}
To make the problem mathematically correct one should follow the same procedure for the force acting on ${\bf R}_{\beta}$ so that the final conditions reads:
\begin{eqnarray}
U_{cg}\frac{\partial f(X_{\alpha},X_{\beta})}{\partial X_{\alpha}}+U_{atom}\frac{\partial g(X_{\alpha},X_{\beta})}{\partial X_{\alpha}}=0 \nonumber \\
U_{cg}\frac{\partial f(X_{\alpha},X_{\beta})}{\partial X_{\beta}}+U_{atom}\frac{\partial g(X_{\alpha},X_{\beta})}{\partial X_{\beta}}=0.
\label{sysdiffeq}
\end{eqnarray} 
This is a system of first order partial differential equations where $g$ and $f$ are the unknown functions in ${\bf \Delta}$ 
and $X_{\alpha}$, $X_{\beta}$ are the variables \cite{note}. Without going into the details of the mathematical properties of such a system, a simple and yet powerful observation clearly shows that a solution may exist only in very special cases but certainly not in general.
This observation is rather simple; a differential equation or a system of differential equations of the first order has got solutions which are uniquely identified by one boundary condition (one for $f$ and one for $g$ in this case). At this point if one goes back to the definition of $f$ and $g$ given in Eqs.\ref{bc1},\ref{bc2}, it is easy to see that in order to have a valid switching function with the correct limiting case at the boundary of ${\bf \Delta}$, there are two boundary conditions for each function, associated to Eq.\ref{sysdiffeq}, to be satisfied: 
\begin{eqnarray}
f(X_{\alpha},X_{\beta})=0; X_{\alpha}=x_{2}, and, X_{\beta}=x_{2}\nonumber\\f(X_{\alpha},X_{\beta})=1; X_{\alpha}=x_{1}, and, X_{\beta}=x_{1}
\label{bc11}
\end{eqnarray}
and 
\begin{eqnarray}
g(X_{\alpha},X_{\beta})=0; X_{\alpha}=x_{1} and, X_{\beta}= x_{1}\nonumber\\g(X_{\alpha},X_{\beta})=1; X_{\alpha}= x_{2}, and, X_{\beta} =x_{2}.
\label{bc22}
\end{eqnarray}
This means the system of equations is overdetermined and a solution in general does not exist . Specifically, if the equation are solved using the condition in $x_{1}$, it may or may not exist a solution such that $f$ in a certain point $x_{2}$ is equal to zero, and equivalently for $g$; of course the same arguments is valid if as a boundary condition is chosen that of $x_{2}$. However even in case a solution exists, there would be no control on the switching region ${\bf \Delta}$ as it is not possible to locate one of the two boundaries {\it a priori}. This aspects makes this approach not convenient for any robust MD algorithm. 
A further point that invalidates the potential approach is the fact that, while ideally $f$ and $g$ should be function solely of $X$, to deal with a simple algorithm, Eq.\ref{sysdiffeq} shows that indeed at least one of the two functions should depend on all the degrees of freedom of the atomistic system, as in the equation the atomistic potential depends on all such degrees of freedom. This may be even possible for simple systems, however as the molecules become larger this approach becomes highly unpractical. One may even think of a more general scheme in the same fashion of what is proposed in Ref.\cite{ensing}, that is to introduce an additional potential $\Phi$ such that the coupling potential reads:
\begin{equation}
U^{coupling}=f(X_{\alpha},X_{\beta})U_{cg}+g(X_{\alpha},X_{\beta})U_{atom}+\Phi
\label{fi}
\end{equation}
where $\Phi$ is equal to zero in ${\bf A}$ and ${\bf B}$ , in order to obtain $U_{cg}$ in ${\bf A}$ and $U_{atom}$ in ${\bf B}$ and it is a certain regular function in ${\bf \Delta}$ such that:
\begin{equation}
{\bf F_{drift}}=\frac{\partial \Phi}{\partial X_{i}};~~i=\alpha, \beta;~~\forall~~X_{\alpha}, X_{\beta} \in {\Delta}. 
\label{eqfidr}
\end{equation}
In this way one obtains a more general  expression for the energy in the switching region and, regarding the forces, the role of $ \frac{\partial \Phi}{\partial X_{i}}$ is that of removing the spurious force due to the switching functions.
At this point one must notice that Eq.\ref{eqfidr} is equivalent to Eq.\ref{partdfiff1} (or Eq.\ref{sysdiffeq}) with the only difference of the presence of $ \frac{\partial \Phi}{\partial X_{i}}$ on the r.h.s.
The conclusions of this work do not change because the problem of the boundary conditions of this differential equation remains the same as for Eq.\ref{partdfiff1} (or Eq.\ref{sysdiffeq}). In this case however there is more flexibility and one can distinguish two situations: (a) $\Phi$ is a known function and $g$ and $f$ unknown; (b) $\Phi$ is unknown and $g$ and $f$ are known.
In case (a) the conclusions drawn before do not change because we still have two first order partial differential equations in $g$ and $f$ and the overdetermination is not removed by the presence of the known term $-\frac{\partial \Phi}{\partial X_{i}}$ on the r.h.s of Eq.\ref{sysdiffeq}. In case (b) we will have again a system of first order partial differential equations where the unknown function is $\Phi$ (Eq.\ref{eqfidr}) and is characterized by two boundary conditions, one in $x_{1}$ and one in $x_{2}$, (i.e. $\Phi=0$), thus the overdetermination is shifted from $f$ and $g$ to $\Phi$.  
\subsection{Conclusions}
We have shown that an adaptive resolution method based on the ansatz of
potential interpolation via switching functions cannot be realized as the mathematical condition of finding a suitable switching function it is likely to have no solution or only trivial ones. It was already shown before that this scheme leads to the violation of Newton third law \cite{pre2,jpa}for the special case $f=1-g$. The arguments presented above add up to the previous one and further show that for the most generic interpolation formula the switching functions do not exist except for some special and trivial cases. In general, for a numerical implementation, such a scheme would not be feasible. This fact does not exclude the possibility that the adaptive resolution can be achieved via other approaches based on the potential energy. In fact, recently Hyden {\it et al.} \cite{hyden} have presented an alternative scheme for adaptive resolution based on potentials. This scheme, rather promising, can be applied also to the quantum-classical interface. However it does not make use of switching functions and looses the numerical simplicity of the interpolation formula together with its physical interpretation which instead is the non trivial advantage of the force based method. In conclusion, the development of adaptive resolution
approaches is a field of rapidly growing interest, the intention of this work is that of fixing some
clear directions along which one can or cannot move in order to develop more
sophisticated and yet numerically simple schemes.

I am grateful to Matej Praprotnik and Kurt Kremer for helpful
comments on the manuscript. This work is supported in part by the
Volkswagen foundation.



\begin{thebibliography}{10}

\bibitem{kax}
G.Lu, E.B.Tadmor, and E.Kaxiras
\newblock Phys.Rev.B {\bf 73}, 024108 (2006).

\bibitem{Rottler:2002}
J.~Rottler, S.~Barsky, and M.~O. Robbins,
\newblock Phys. Rev. Lett. {\bf 89}, 148304 (2002).

\bibitem{Csanyi:2004}
G.~Csanyi, T.~Albaret, M.~C. Payne, and A.~D. Vita,
\newblock Phys. Rev. Lett. {\bf 93}, 175503 (2004).

\bibitem{Laio:2002}
A.~Laio, J.~VandeVondele, and U.~R{\" o}thlisberger,
\newblock J. Chem. Phys. {\bf 116}, 6941 (2002).

\bibitem{Jiang:2004}
D.~E. Jiang and E.~A. Carter,
\newblock Acta Materialia {\bf 52}, 4801 (2004).

\bibitem{Lu:2005}
G.~Lu and E.~Kaxiras,
\newblock Phys. Rev. Lett. {\bf 94}, 155501 (2005).

\bibitem{jcp}
M.~Praprotnik, L.~Delle Site, and K.~Kremer,
\newblock J. Chem. Phys. {\bf 123}, 224106 (2005).

\bibitem{pre1}
M.~Praprotnik, L.~Delle Site, and K.~Kremer,
\newblock Phys.Rev.E {\bf 73}, 066701 (2006).

\bibitem{ensing}
B.Ensing, S.O.Nielsen, P.B.Moore, M.L.Klein, and M.Parrinello
\newblock J.Chem.Th.Comp.{\bf 3} 1100 (2007).


\bibitem{pre2}
M.~Praprotnik, K.~Kremer, and L.~Delle Site
\newblock Phys. Rev. E  {\bf 75}, 017701 (2007).

\bibitem{jpa}
M.~Praprotnik, K.~Kremer, and L.~Delle Site
\newblock J.Phys.A:Math.Theor. {\bf 40}, F281 (2007).

\bibitem{jcp2}
M.~Praprotnik, L.~Delle Site, and K.~Kremer,
\newblock J. Chem. Phys. {\bf 126}, 134902 (2007).

\bibitem{jpcm}
M.~Praprotnik, S.Matysiak, L.~Delle Site, K.~Kremer, and C.Clementi
\newblock J.Phys:Cond.Matt. {\bf 19}, 292201 (2007).

\bibitem{note}
It must be notice that when a molecule in  ${\bf \Delta}$ (for example molecule $\alpha$) interacts with a molecule in ${\bf B}$ (molecule $\beta$), $f$ (or equivalently $g$) should take the form $f(X_{\alpha},X_{\beta})=f(X_{\alpha},x_{2}), \forall X_{\beta} \ge x_{2}$. This translates to the fact that the level of resolution in the interaction is decided solely by the molecule in ${\bf \Delta}$, while the resolution of molecule $\beta$ is the same in all region ${\bf B}$. This, obviously, is not the case if $\beta$ is in ${\bf A}$, as the lower resolution always determines the kind of interaction. This latter case naturally comes from the definition of $f$ and $g$ as underlined before, while the former case is an obvious extension of such a definition; in any case it does not enter and it is not relevant for the conclusions of this work and it is given here only for the sake of clarity.
\bibitem{hyden}
A.Heyden, H.Lin, and D.G.Truhlar
\newblock J.Phys.Chem.B {\bf 111}, 2231 (2007).
\end{thebibliography}
\end{document}